\documentclass[sigconf]{acmart}
\AtBeginDocument{%
  \providecommand\BibTeX{{%
    \normalfont B\kern-0.5em{\scshape i\kern-0.25em b}\kern-0.8em\TeX}}}

\copyrightyear{2025}
\acmYear{2025}
\setcopyright{rightsretained}
\acmConference[CHI EA '25]{Extended Abstracts of the CHI Conference on
Human Factors in Computing Systems}{April 26-May 1, 2025}{Yokohama, Japan}
\acmBooktitle{Extended Abstracts of the CHI Conference on Human Factors in
Computing Systems (CHI EA '25), April 26-May 1, 2025, Yokohama,
Japan}\acmDOI{10.1145/3706599.3719736}
\begin{document}





\title{Empowering Social Service with AI: Insights from a Participatory Design Study with Practitioners}


\author{Yugin Tan}
\email{tan.yugin@u.nus.edu}
\affiliation{
  \institution{School of Computing}
  \institution{National University of Singapore}
  \country{Singapore}
}

\author{Soh Kai Xin}
\email{kaisoh2030@u.northwestern.edu}
\affiliation{
  \institution{School of Communications}
  \institution{Northwestern University}
  \country{United States of America}
}

\author{Zhang Renwen}
\email{r.zhang@nus.edu.sg}
\affiliation{
  \institution{Department of Communications and New Media}
  \institution{National University of Singapore}
  \country{Singapore}
}
\author{Lee Jungup}
\email{swklj@nus.edu.sg}
\affiliation{
  \institution{Department of Social Work}
  \institution{National University of Singapore}
  \country{Singapore}
}

\author{Meng Han}
\email{han.meng@u.nus.edu}
\affiliation{
  \institution{School of Computing}
  \institution{National University of Singapore}
  \country{Singapore}
}

\author{Biswadeep Sen}
\email{e0989386@u.nus.edu}
\affiliation{
  \institution{School of Computing}
  \institution{National University of Singapore}
  \country{Singapore}
}

\author{Lee Yi-Chieh}
\email{yclee@nus.edu.sg}
\affiliation{
  \institution{School of Computing}
  \institution{National University of Singapore}
  \country{Singapore}
}

\renewcommand{\shortauthors}{Tan et al.}

\begin{abstract}
In social service, administrative burdens and decision-making challenges often hinder practitioners from performing effective casework. Generative AI (GenAI) offers significant potential to streamline these tasks, yet exacerbates concerns about overreliance, algorithmic bias, and loss of identity within the profession. We explore these issues through a two-stage participatory design study. We conducted formative co-design workshops (\textit{n=27}) to create a prototype GenAI tool, followed by contextual inquiry sessions with practitioners (\textit{n=24}) using the tool with real case data. We reveal opportunities for AI integration in documentation, assessment, and worker supervision, while highlighting risks related to GenAI limitations, skill retention, and client safety. Drawing comparisons with GenAI tools in other fields, we discuss design and usage guidelines for such tools in social service practice.
\end{abstract}




\maketitle

\section{Introduction}

The social service sector is crucial for a just and humane society, addressing various social injustices and supporting the well-being of individuals and communities \cite{difranks2008social}. Social service agencies render critical services such as client home visits, case analysis, and intervention planning, aiming to aid vulnerable clients in the best way possible. Social service practitioners (SSPs) must consider the myriad interacting factors in clients' social networks, physical environment, and intrapersonal cognitive and emotional systems, in the context of clients' beliefs, perceptions, and desires, all while respecting individual worth and dignity \cite{rooney2017direct}. This requires extensive training and hands-on experience, which may be challenging given the endlessly varying situations a worker may encounter \cite{rooney2017direct}.

These demands put a strain on the limited pool of social service manpower \cite{cho2017determinants}. SSPs grapple with time-consuming data processing \cite{singer_ai_2023, tiah_can_2024} and administrative \cite{meilvang_working_2023} tasks, compounded by the psychologically stressful nature of the job \cite{kalliath2012work}. Newer workers may experience challenging, unfamiliar client interactions, while experienced workers also face the added burden of mentoring junior workers. Current artificial intelligence (AI) systems aim to alleviate this through \textit{decision-support tools} \cite{gambrill2001need, de2020case, brown2019toward} which provide statistically validated assessments of client conditions \cite{gillingham2019can, van2017predicting}, thereby raising service quality and consistency. However, attempts to integrate AI systems in social service have faced a myriad of difficulties: mistrust in and uncertainty about technology \cite{gambrill2001need, brown2019toward}, worker confusion due to unclear organisational direction \cite{kawakami2022improving}, AI systems' inability to include local and contextual knowledge, and fear of AI potentially replacing jobs. As such, many attempts have been met with failure and inadequacies \cite{saxena2021framework, kawakami2022improving}.

Recent technological advancements, in particular generative AI (GenAI), have the potential to play a greater, transformative role in the social service sector. GenAI systems are capable of a wider range of tasks, including analysing written case recordings, performing qualitative risk assessments, providing crisis assistance, and aiding prevention efforts \cite{reamer_artificial_2023}.  However, despite their advantages, these systems may exacerbate existing concerns with AI, risk overreliance \cite{van2023chatgpt} and potentially erode the human skills and values \cite{littlechild2008child, oak2016minority} core to the social service profession. 
As a novel technology, the efficacy of GenAI systems are still largely under-explored and untested in extant research \cite{gambrill2001need, de2020case, brown2019toward}, thus necessitating deeper scrutiny of AI technologies in social service. We therefore ask, \textbf{RQ1:} How can GenAI help SSPs in their daily work, and \textbf{RQ2:} How can GenAI help SSPs at an organisational level?

To address this gap, we conducted a two-stage study with 51 SSPs, to better understand practitioners' perspectives on implementing AI in the social service sector. 
We first ran preliminary workshops with 27 SSPs to co-design a prototype GenAI tool, then conducted user testing and focus group discussions (FGDs) with 24 additional SSPs. Through this study, we offer an empirically-grounded understanding of how workers use and perceive AI, shedding new light on the challenges and opportunities of GenAI in the social service sector and beyond. We conclude with suggestions for social service organisations and design and usage guidelines for AI tools in the sector.

\section{Related Work}
\label{sec:relatedwork}

\subsection{Current AI Tools for Social Service Practitioners}
\label{subsec:relatedtools}

Artificial Intelligence (AI) has long been used for risk assessments, decision-making, and workload management in sectors like child protection services and mental health treatment \cite{fluke_artificial_1989, patterson_application_1999}. 
Recent applications in clinical social work include risk assessments \cite{gillingham2019can, jacobi_functions_2023, liedgren_use_2016, molala_social_2023}, public health initiatives \cite{rice_piloting_2018}, and education and training for practitioners \cite{asakura_call_2020, tambe_artificial_2018}. Present studies on case management focus mainly on decision support tools \cite{james_algorithmic_2023, kawakami2022improving}, especially predictive risk models (PRMs) used to predict social service risks and outcomes \cite{gillingham2019can, van2017predicting}. A prominent example is the Allegheny Family Screening Tool (AFST), which assesses child abuse risk using data from US public systems \cite{chouldechova_case_2018, vaithianathan2017developing}. Elsewhere, researchers have also piloted AI systems to predict social service needs for the homeless using Medicaid data \cite{erickson_automatic_2018, pourat_easy_2023} or promote health interventions like HIV testing among at-risk populations \cite{rice_piloting_2018, yadav_maximizing_2017}.

Beyond such tools, however, the sector also stands to benefit immensely from newer forms of AI, such as GenAI. SSPs work in time-poor environments \cite{tiah_can_2024}, being often overwhelmed with tedious administrative work \cite{meilvang_working_2023} and large amounts of paperwork and data processing \cite{singer_ai_2023, tiah_can_2024}. GenAI is well placed to streamline and automate tasks such as the formatting of case notes, the formulation of treatment plans, and the writing of progress reports, allowing valuable time to be spent on more meaningful work, such as client engagement and the improvement of service quality \cite{fernando_integration_2023, perron_generative_2023, tiah_can_2024, thesocialworkaimentor_ai_nodate}. There is, however, scant research on GenAI in the social service sector \cite{wykman_artificial_2023}.





\subsection{Challenges in AI Use in Social Service}
\label{subsec:relatedworkaiuse}

Despite its evident benefits, multiple challenges plague the integration of AI and its vast potential into real-life social service practice.
When employing algorithmic decision-making systems, practitioners often experience tension in weighing AI suggestions against their own judgement \cite{kawakami2022improving, saxena2021framework}, being uncertain of how far they should rely on the machine. 
Workers are often reluctant to fully embrace AI assessments due to its inability to adequately account for the full context of a case \cite{kawakami2022improving, gambrill2001need}, and lack of clarity and transparency on AI systems and limitations \cite{kawakami2022improving}. Brown et al. \cite{brown2019toward} conducted workshops using hypothetical algorithmic tools 
and found similar issues with mistrust and perceived unreliability. Furthermore, introducing AI tools can create new problems of its own, causing confusion and distrust amongst workers \cite{kawakami2022improving}. Such factors are critical barriers to the acceptance and effective use of AI in the sector.

\citeauthor{meilvang_working_2023} (2023) cites the concept of \textit{boundary work}, which explores the delineation between "monotonous" administrative labour and "professional", "knowledge-based" work drawing on core competencies of SSPs. While computers have long been used for bureaucratic tasks such as client registration, the introduction of decision support systems like PRMs stirred debate over AI "threatening professional discretion and, as such, the profession itself" \cite{meilvang_working_2023}. Such latent concerns arguably drive the resistance to technology adoption described above. GenAI is only set to further push this boundary, 
with its ability to formulate detailed reports and assessments that encroach upon the "core" work of SSPs.
Introducing these systems exacerbates previously-raised issues such as understanding the limitations and possibilities of AI systems \cite{kawakami2022improving} and risk of overreliance on AI \cite{van2023chatgpt}, and requires a re-examination of where users fall on the algorithmic aversion-bias scale \cite{brown2019toward} and how they detect and react to algorithmic failings \cite{de2020case}. We address these critical issues through an empirical, on-the-ground study that to our knowledge is the first of its kind since the new wave of GenAI.

\section{Study Design}
\label{sec:stage1design}

We engaged in a two-stage study: a formative, participatory design study to understand the opportunities and challenges perceived by our participants and to co-design a GenAI tool, followed by an evaluative contextual inquiry to assess the effectiveness of the resulting tool. We partnered with two local, government-funded social service agencies (SSAs) in a Southeast Asian country that had expressed interest in adopting GenAI. Agency A focused on family-oriented casework, handling mostly walk-in clients and taking them through the full process \cite{rooney2017direct} of exploration, assessment, implementation, and eventually termination. Agency B worked more with schools and youths, partnering with education institutes to render assistance to children or teenagers in need. Both agencies used English as a working language and for all official documentation. In a small proportion of cases, Agency A's client interactions took place in a different language that was more comfortable for the client; in these instances, workers would either record their notes in English or manually translate them back into English before taking them back to the agency for further work. All participant interactions in this study were also conducted in English.


In \textbf{stage 1, the co-design phase}, we conducted two 90-minute long workshops with agencies A and B in November 2023. A total of 27 SSPs were involved, hailing from different roles and experience levels\footnote{Pictures of the workshops are in Appendix \ref{appendix:workshops}.}. We aimed to understand 1) the nature of the day-to-day work that our participants performed and what opportunities they perceived for using AI to help with it, and 2) the perceived risks and challenges of AI use to shape the design of the second phase of our study. In each workshop, we briefly introduced LLMs in the form of ChatGPT\footnote{At the time of the workshops (October 2023), ChatGPT (GPT-3.5) was the most well-known LLM.}, then conducted brainstorming and sketching sessions in small groups of 4-6. Full details of the workshops are in Appendix \ref{appendix:workshops}.

\begin{figure}
\centering
\includegraphics[width=2.15in]{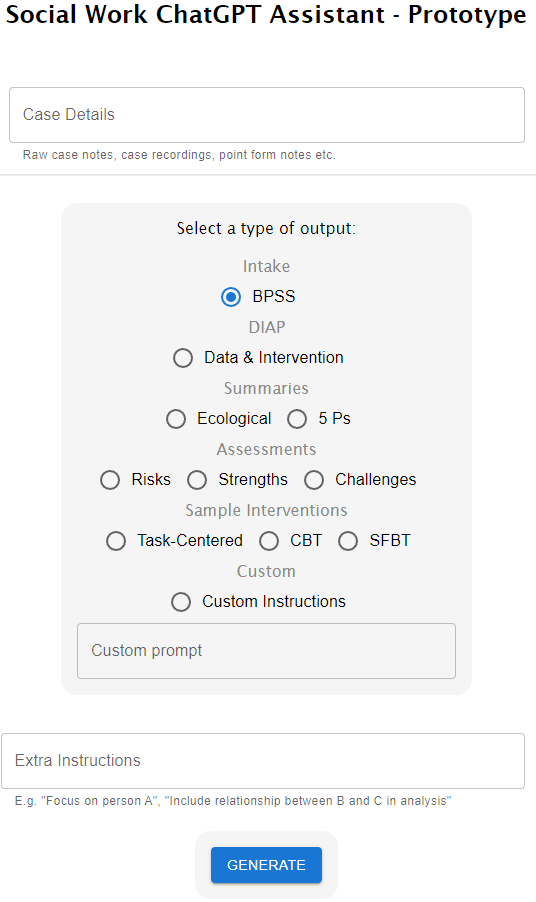}
\caption{Prototype AI Tool}
\Description{A screenshot of our prototype AI tool. It has a text box for entry at the top, a list of radio buttons to select output modalities, a text box for extra instructions to the model a user might want to input, and a "generate" button.}
\label{fig:participantsAndTool}
\end{figure}

Based on the opportunities and concerns identified in the workshops (see Section \ref{findings:workshop} or Appendix \ref{appendix:workshopFindings}), we created a prototype assistant tool (Fig. \ref{fig:participantsAndTool}, right). The tool comprised an input text box for users to enter details about their client and a number of different types of output options. Users could select an output option based on a desired use case, then click a "generate" button to produce an LLM-written\footnote{OpenAI GPT-4 Turbo, gpt-4-1025-preview.} response. These output options addressed various manual and mental labour difficulties in social work, and simulated potential uses addressing these difficulties. In response to complaints about manual work, the tool offered options to rewrite workers' rough notes into various organization-wide formats (e.g. BPSS, DIAP, Ecological, 5Ps). Users also raised points about assessments, case conceptualization, and ideation. We included options for strength, risk, and challenge assessments, following common social service practices \cite{rooney2017direct}. Finally, we added options to generate client intervention plans according to three common theoretical models: CBT, SFBT\footnote{CBT: Cognitive Behavioural Therapy; SFBT: Solution-Focused Brief Therapy.}, and Task-Centred Interventions.

In \textbf{stage 2, we conducted focus group discussions} (FGDs), where we simulated a contextual inquiry process \cite{holtzblatt2017contextual} by asking participants to walk us through their use of our tool and explain their thought processes along the way. We also encouraged them to point out weaknesses or flaws in the system and suggest potential improvements. We conducted the FGDs in groups of 2-4 SSPs from Agency A (the family service centre agency), with a mix of workers of varying seniority levels. We held a total of 8 sessions totalling 24 participants (Fig. \ref{fig:participantsAndTool}, left). Each session averaged 45-60 minutes in length and was audio-recorded for transcription and analysis. These sessions were conducted in May 2024.

In these sessions, we sought to understand the \textit{opportunities} of AI use in social work. Participants brought along different types of anonymised case files (e.g., rough short-hand notes, complete reports, intake files) from recent clients they had worked with. We explained the various functions of the system, then told participants to imagine themselves using it to help with the cases they had on hand, exploring a range of use cases from generating documentation reports to planning future sessions or interventions for the client. We also identified early on that case supervision by senior workers was a key means of addressing the difficulties faced by junior workers and encouraging worker development. We therefore sought to understand how AI could play a collaborative role in the supervision process. In sessions with supervisors present, we asked these more senior workers how they felt the various functions of the system could assist them in their discussion of cases with their supervisees. 

We also aimed to understand the possible \textit{challenges} of using AI tools in social work. We asked participants to review the quality of the tool's outputs, compare them to their own, and identify areas where they would fall short of expectations or fail to be useful in their daily work. We note that while our platform is based on GPT-4, a general purpose model not tuned for social work analysis\footnote{Specialised AI systems for social work case management do exist (e.g. \cite{socialworkmagic2024, caseworthy2024}, but these are focused on user experience and do not provide greater insight than vanilla GPT; furthermore, \cite{socialworkmagic2024} advises users to take its assessments and suggestions as only a starting point, which aligns with how we position our tool.}, our aim was not to discover the specific weaknesses of GPT-4 or any LLM in particular. Rather, it was to understand what participants perceived to be good outputs from an AI system, and in the process understand how they might be affected by any possible shortcomings of such a system. 

Finally, we explored some of the longer-term effects of AI use, such as potential overreliance on the tool and the possibility of becoming overly trusting of the system's output. Senior workers in particular were asked about how they perceived junior workers new to the sector having such a system to help them with their daily work.

We performed qualitative thematic analysis \cite{braun2006using} on the transcripts, adopting a bottom-up, inductive approach to data coding. This process is detailed in Appendix \ref{appendix:analysis}, with the findings presented in the next section.

\section{Findings}
\label{sec:stage2findings}

\subsection{Findings from Stage 1: Workshop and Co-Design}
\label{findings:workshop}

Through the workshops, participants revealed a few major aspects of their work that could be assisted by AI. For brevity, detailed findings are in Appendix \ref{appendix:workshopFindings}. In brief, the main findings are below.

Documentation was a major pain point, with our participants needing to document "anything and everything", writing systematic reports in different structured formats (e.g. bio-psychological scales, risk factor assessments) or repacking the same content for different stakeholders like colleagues or other agencies, creating tedious duplicate work. Participants hence expressed a desire for a tool to help with manual labour, like turning point-form notes into formal reports, in various formats such as the 5Ps, DIAP, or BPSS\footnote{5Ps: Presenting problem, Predisposing factors, Precipitating factors, Perpetuating factors, Protective factors; DIAP: Data, Intervention, Assessment, Plan; BPSS: Bio-Psycho-Social-Spiritual}. Some senior personnel also suggested that AI could help workers incorporate theoretical concepts into their written work.

\begin{table}
\centering
\label{tab:agencyA}

\begin{tabular}{ll|ll}
\toprule
\multicolumn{2}{c|}{\textbf{Centre 1}} & \multicolumn{2}{c}{\textbf{Centre 2}} \\
\midrule
Code & Role        & Code & Role        \\
\midrule
W1   & Social Worker & W9   & Social Worker  \\
W2   & Social Worker & W10  & Social Worker  \\
W3   & Social Worker & C1   & Snr Counsellor \\
\midrule
S1   & Supervisor    & S7   & Senior SW      \\
S2   & Supervisor    & W7   & Social Worker  \\
S3   & Supervisor    & W8   & Social Worker  \\
\midrule
      &              & S6   & Senior SW      \\
      &              & W5   & Social Worker  \\
\bottomrule
\multicolumn{2}{c|}{\textbf{Centre 3}} & \multicolumn{2}{c}{\textbf{Centre 4}} \\
\midrule
Code & Role        & Code & Role         \\
\midrule
S4   & Supervisor  & S8   & Senior SW    \\
S5   & Supervisor  & S9   & Senior SW    \\
W4   & Social Worker & S10  & Senior SW   \\
\midrule
      &             & W11  & Social Worker \\
      &             & W12  & Social Worker \\
      &             & W13  & Social Worker \\
\bottomrule
\end{tabular}

\caption{Focus Group Participants from Agency A}
\end{table}

Participants also noted challenges with case formulation, in which piecing together disparate information across many pages of case notes to craft and justify an assessment is cognitively challenging and time-consuming, particularly during direct interaction with a client. This may inadvertently cause them to miss certain key insights or red flags only evident from looking at the gathered information as a whole. Participants therefore also suggested systems for generating case assessments, to improve output quality and adherence to industry-standard terms, and intervention planning, to generate possible plans for helping the client that the practitioner could consider and choose from.




Finally, participants also noted a few potential areas of risk. Privacy of client data was unilaterally mentioned by all, a universal concern core to the social sector. The storage of personal client information was a significant concern, particularly regarding the possibility of workers mistakenly entering sensitive information into the system. Regarding staff competency, a common concern was the impact of AI taking on an increasing part of the worker's job scopes. Specifically for analytical or ideative tasks, some seniors were concerned about the loss of critical thinking skills of junior workers who might become overreliant on the tool to perform their work for them. Multiple participants also raised the possibility of inaccurate outputs from an AI system, particularly risky when less experienced workers fail to tell when the AI's output might be suboptimal and proceed to adopt its suggestions anyway. 

\subsection{Findings from Stage 2: Focus Group Discussions}

\subsubsection{Documentation} 
\label{subsubsec:discussionuses}

Participants found many applications of GenAI in helping with multiple writing-focused tasks in the social service sector, such as summarising intake information, formatting case recordings, and writing reports. Participants generally were happy to embrace AI for such purposes; for documentation tasks such as writing case reports, the tool's outputs were largely in line with what they required, allowing them to simply "copy and paste" (W1, W2, W13) the outputs for direct use. This was, in a large way, down to how much of our participants' regular daily work focused on consistently structured, fixed-format reports. One strength evident in GPT-4 and our prototype was its ability to consistently follow instructions to produce outputs in a desired format, such as the 5Ps format (S1, S3, W1). W8, for instance, quoted, "\textit{being new, it really helps in categorising these items... I like the fact that it segregates all the [different categories]}".

Even when the output was imperfect, participants often expressed a willingness to work around these errors and make manual corrections where needed. For instance, W2 suggested they would "sift through and pick out" the more relevant parts of an overly lengthy report, while W1 would "\textit{copy and paste, then amend if I need to amend}" when the tool misinterpreted a nickname given by the worker to the client.


\subsubsection{Brainstorming}

Workers felt that the structured and detailed outputs of the tool encouraged and facilitated their cognitive processes in analysing a case. W8 felt the way the tool categorised the issues in their client's case pushed them to "think more" about how they viewed it. In a similar vein, looking at a generated CBT assessment prompted W5 to consider "\textit{certain things also that I should probably look into}", which they might otherwise have overlooked. W11 commented on how the "very in-depth" explanations given by the tool helped them "expand on what they already have". Even a senior worker, S4, called the tool's analysis of a case "\textit{really helpful - it's expanding my perspective already}".

Junior workers in particular appreciated the guidance from the tool, especially with tasks they were less familiar with. For instance, W7 had not attended formal CBT training, and thus felt the tool gave them "some idea where to start" in formulating a CBT intervention plan for their client. W4 also liked having sample interventions from the tool; being a newer worker, they were uncertain of which plan of action to take, so the tool gave them a "better understanding" of the different ways to move forward with the case. Meanwhile, more generally, the tool helped with guiding workers towards formulating a course of action, such as by suggesting interventions they might not have thought of (W4) and thereby prompting newer staff with a "direction" to work towards (W5), or by being "very useful" in helping new workers prepare for sessions with clients (W9).

Supervisors, too, agreed that the tool was useful for junior workers. S4 reiterated that it could "expand the worker's perspective", and C1 called it "a good start [to help] staff think about" the case if they "got stuck" with something. S6 cited the example of how an SFBT output provided a "really good foundation" for questions for workers to ask their clients. S6 also felt the tool provided a good framework and guideline, suggesting it as a way to "polish" newer workers' skills: "\textit{For all those who are really new and do not really know how to formulate interventions or theory support and all that. I think it's quite useful... or, if they are really lost, then they can probably try the different things that are written here}." 

\subsubsection{Supervision}

The use of AI to aid in supervision emerged as a key theme. Supervision sessions consist of junior workers discussing cases with their supervisors, to refine and improve their assessment of the client. Given this, the ability of AI to quickly generate lists of ideas provided useful starting points for discussion and reflection with supervisees (C1). Many (W4, W5) suggested that the tool provided useful intervention suggestions so that workers could "ask their supervisors like, maybe, you know, maybe I can try this" (S6). From the supervisors' perspectives, the tool helped to improve and expedite the supervision process by prompting them with questions to ask supervisees: "I can bring [this list of exception questions\footnote{Asking "exception questions" to clients is a technique used in Solution-Focused Brief Therapy.}] to supervision to see whether my supervisee has used these questions... I can ask my supervisee, okay, if you have to ask this exception question to the client, how comfortable do you feel? So we can have that discussion" (C1). S4 and S7, meanwhile, highlighted the AI's ability to quickly "concretise theoretical models" to build on during sessions. 

Some even suggested that the tool could itself serve as a supervisor to help newer workers. W9 called the tool a "readily-available supervisor to get us thinking," referencing how their supervisors frequently prompted them with questions to think about their case more. W4 meanwhile felt the tool could help them move forward with a case "without having to consult with their supervisor", in instances where they were uncertain of how to proceed.

\subsubsection{Concerns and Issues}

Participants raised a few issues with using AI in their work. Agency A's focus on providing family service meant that they prioritised addressing safety and risk concerns (W2), such as possible self-harm, suicide, or harm towards others. Participants were "very particular about... risks" (S9), treating it as their top priority (W1, S10) and "at the front" for all client interventions and risk assessments (S2, S9). Preventing imminent physical and psychological harms such as incarceration, abuse, and addiction were therefore cited as "non-negotiables" (W4) and primary risk factors (W2). Thus, they expressed concern when the tool "didn't exactly highlight" (W2) or entirely omitted (W1) safety risks in its output, such as in assessing a case of intra-family conflict: "\textit{The [risk of] violence is not highlighted. Where is the violence?}" (W1).

There were also worries over overuse and overreliance on AI. Participants were divided on this issue. The centre director (D1) quoted, "\textit{One of our concerns is... [will using this AI] actually disable our ability to make assessments?}" S3 agreed about the risk of over-reliance by "spoonfeeding" junior workers, and S9 noted the need for workers to "\textit{still use their brain... otherwise... they just rely on this}." S9 also worried about how junior workers might fail to recognise suboptimal AI outputs and be misled as a result. As a result, some emphasised the need for a balance between AI use and human intervention. S10 remarked, \textit{"It's the supervisor's role to keep emphasising to the supervisee, to not be married to this assessment"}, and that the AI's outputs were often "just guidelines" rather than a gospel to be followed. S9 agreed that "\textit{the expectation [for use] needs to be very clearly communicated}", highlighting the need for careful and judicious implementation of any AI system. 

However, others felt this to not be a major problem, due to the inherent focus of the profession on face-to-face client interactions. On practitioners blindly following AI recommendations, S7 commented, "\textit{it will eventually be clear that this was the recommendation of [the tool] but then you went and did something else... and then you'll see that it's just a mess in that way.}" S3 agreed: "\textit{When they do their work, it is mostly real-time. You don't go back to pen and paper and start to develop [a solution], but it's more about what is presented to them immediately [and how they react].}"

Finally, there were multiple instances where participants found the AI's output to be inadequate. These often centred around the tool failing to recognise more subtle factors in a given case that would affect the preferred course of action. For example, when the system recommended a family therapy workshop, W5 recognised that the reluctance of their client to embrace outside help made such an intervention unsuitable. In another case, when the tool recommended CBT to aid a client, W11 judged that their client lacked the intellectual capacity for such therapy to help. Knowledge of local contextual factors was another common point. S6, for instance, commented on how "\textit{there are a few parenting workshops available [here in our country] and [we know] how they are structured}," so they would be able to assess which options were suitable for their client in a way that the AI, lacking local knowledge, would not. W5 also commented on the importance of practitioners' instincts and experience: \textit{"Many of us who have been in the sector before, probably were like, this won't work."}

\section{Discussion and Conclusion}


In this study, we uncovered multiple ways in which GenAI can be used in social service practice. While some concerns did arise, practitioners by and large seemed optimistic about the possibilities of such tools, and that these issues could be overcome. We note that while most participants found the tool useful, it was far from perfect in its outputs. This is not surprising, since it was powered by a generic LLM rather than one fine-tuned for social service case management. However, despite these inadequacies, our participants still found many uses for most of the tool's outputs. Many flaws pointed out by our participants related to highly contextualised, local knowledge. To tune an AI system for this would require large amounts of case files as training data; given the privacy concerns associated with using client data, this seems unlikely to happen in the near future. What our study shows, however, is that GenAI systems need not aim to be perfect to be useful to social service practitioners, and can instead serve as a complement to the critical "human touch" in social service.

We draw both inspiration and comparisons with prior work on AI in other settings. Studies on creative writing tools showed how the "uncertainty" \cite{wan2024felt} and "randomness" \cite{clark2018creative} of AI outputs aid creativity. Given the promise that our tool shows in aiding brainstorming and discussion, future social service studies could consider AI tools explicitly geared towards creativity - for instance, providing side-by-side displays of how a given case would fit into different theoretical frameworks, prompting users to compare, contrast, and adopt the best of each framework; or allowing users to play around with combining different intervention modalities to generate eclectic (i.e. multi-modal) interventions.

At the same time, the concept of supervision creates a different interaction paradigm to other uses of AI in brainstorming. Past work (e.g. \cite{shaer2024ai}) has explored the use of GenAI for ideation during brainstorming sessions, wherein all users present discuss the ideas generated by the system. With supervision in social service practice, however, there is a marked information and role asymmetry: supervisors may not have had the time to fully read up on their supervisee's case beforehand, yet have to provide guidance and help to the latter. We suggest that GenAI can serve a dual purpose of bringing supervisors up to speed quickly by summarising their supervisee's case data, while simultaneously generating a list of discussion and talking points that can improve the quality of supervision. Generalising, this interaction paradigm has promise in many other areas: senior doctors reviewing medical procedures with newer ones \cite{snowdon2017does} could use GenAI to generate questions about critical parts of a procedure to ask the latter, confirming they have been correctly understood or executed; game studio directors could quickly summarise key developmental pipeline concerns to raise at meetings and ensure the team is on track; even in academia, advisors involved in rather too many projects to keep track of could quickly summarise each graduate student's projects and identify potential concerns to address at their next meeting.

In closing, we are optimistic about the potential for GenAI to significantly enhance social service practice and the quality of care to clients. Future studies could focus on 1) longitudinal investigations into the long-term impact of GenAI on practitioner skills, client outcomes, and organisational workflows, and 2) optimising workflows to best integrate GenAI into casework and supervision, understanding where best to harness the speed and creativity of such systems in harmony with the experience and skills of practitioners at all levels.

\begin{acks}
The authors would like to thank the directors and workers at Agency A for our close working relationship over the past 18 months. 

We also acknowledge the use of GPT-4o for generating ideas for the \textit{paper title}, giving suggestions for the \textit{abstract} after the rest of the paper was completed, and \textit{shortening} parts of the paper after they were written.

This work was supported by the National University of Singapore's Centre for Computational Social Science and Humanities (CSSH), under fund number A-8002954-01-00.
\end{acks}

\bibliographystyle{ACM-Reference-Format}
\bibliography{references}

\appendix
\newpage

\section{Workshop Images}
\label{appendix:workshops}
\begin{figure}[H]
    \centering
    \includegraphics[scale=0.45]{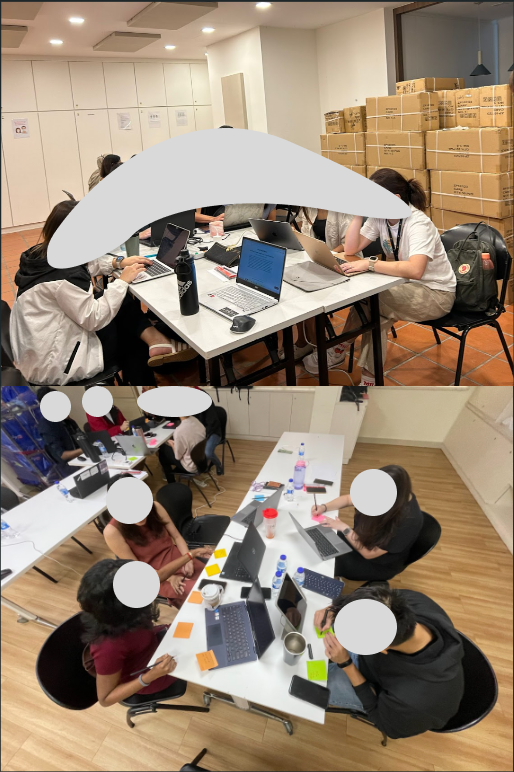}
    \caption{Workshop participants}
    \label{fig:workshopParticipants}
    \Description{A picture of workshop participants.}
\end{figure}

\begin{figure}
    \centering
    \includegraphics[scale=0.25]{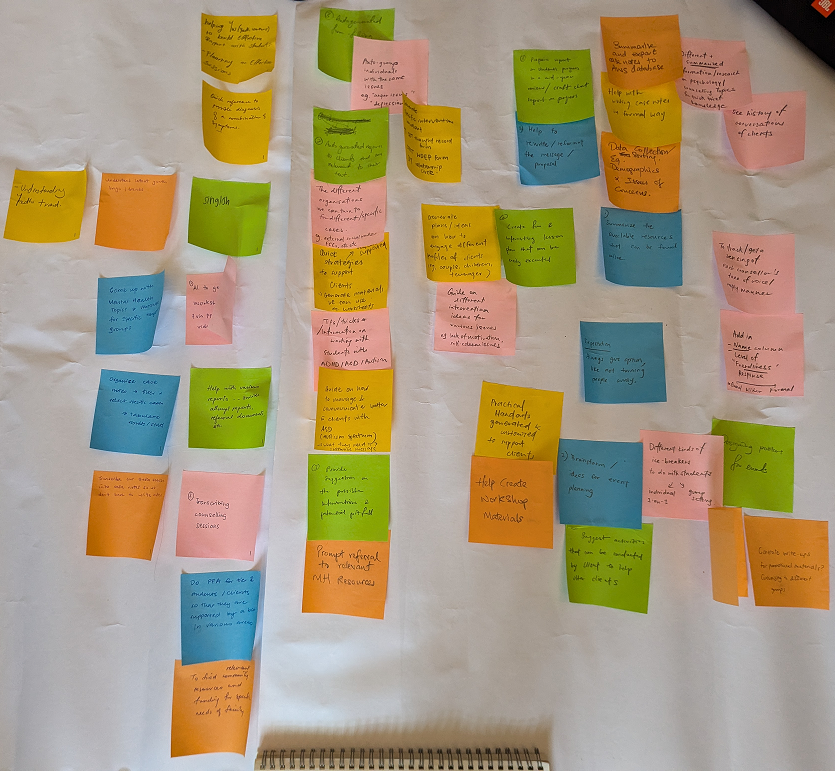}
    \caption{Notes generated by participants}
    \label{fig:workshopPostIts}
    \Description{Post-It notes generated by participants.}
\end{figure}

\section{Workshop Design}

We began with two workshops (Fig. \ref{fig:workshops}) with two social service agencies, A and B, in November 2023. This provided a high-level understanding of the social service sector pertaining to possible AI solutions and possible associated implementation challenges. A total of 27 SSPs were involved, hailing from different roles and experience levels (Table \ref{tab:workshopParticipants}),
giving us a wide range of perspectives to draw upon. 

Each workshop lasted around 90 minutes, comprising a briefing and discussion session. In the briefing, we introduced LLMs in the form of ChatGPT\footnote{At the time of the workshops (October 2023), ChatGPT was the main LLM widely accessible to the public.}. We demonstrated AI's strengths (e.g., brainstorming, ideation \cite{gero2023social, shaer2024ai}, summarisation \cite{wang2023chatgpt}, formatting and rewriting of text \cite{bhattaru2024revolutionizing}) and weaknesses, including hallucination \cite{maynez2020faithfulness} and limited context windows \cite{liu2024lost}. This ensured that all participants, including those with only a cursory understanding or awareness of ChatGPT\footnote{ChatGPT had been available for around 10 months at this point. Some workers had tried using it, even experimenting with it for their work, while others had never used it themselves.} had a baseline understanding of LLMs.

Next, we split participants into tables of 4-6 people each for small-group discussions. Here, we sought to encourage participants to freely express their thoughts and build on one another's ideas \cite{bonnardel2020brainstorming}. Participants were given stacks of Post-Its and asked a series of questions. For each question, they spent around five minutes writing their responses on individual Post-Its. Then, they pasted their Post-Its on a board and clustered similar ideas together, allowing common themes to emerge. To kickstart the discussion with more familiar topics, we asked the participants about their work: \textit{"What does a day at work look like for you?"}, \textit{"What difficulties have you recently faced in your work?"}. We then brought AI into the picture:\textit{ "How can AI help you in your work?"}, \textit{"What expectations do you have for an AI-powered social service tool?"}. Having elicited some ideas, we then introduced two rounds of 6-8-5 sketching where participants quickly drew out concepts of possible AI tools that could help them. Finally, we asked, \textit{"What are some concerns you might have with using AI in your work?"}.

After the workshops, the researchers transferred the physical Post-Its to a Miro board for further analysis and deduced themes from the clustering done during the workshops.

\begin{table}[H]
\centering
\begin{tabular}{cc|cc}
    \toprule
    \multicolumn{2}{c|}{\textbf{Agency A}}&\multicolumn{2}{c}{\textbf{Agency B}} \\
    \midrule
    Code&Role&Code&Role\\
    \midrule
    TS1 & Snr Social Worker & CD1 & Director\\
    TS2 & Social Worker & CY1 & Youth Work \\
    TS3 & Social Worker & CY2 & Youth Work \\ 
    TS4 & Social Worker & CY3 & Youth Work \\
    TS5 & Social Worker & CP1 & Youth Projects\\
    TS6 & Social Worker & CP2 & Youth Projects\\
    TS7 & Social Worker & CP3 & Youth Projects\\
    TS8 & Social Worker & CP4 & Youth Projects \\
    TS9 & Social Worker & CC1 & Counsellor \\
    TS10 & Social Worker & CC2 & Counsellor \\
    TD1 & Exec. Director & CC3 & Counsellor\\
    TD2 & Snr Director & CC4 & Counsellor\\
    TP1 &  Psychologist & CC3 & Counsellor\\
    && CC4 & Counsellor\\
\bottomrule   
\end{tabular}
\caption{Workshop Participants from Agency A and Agency B}
\label{tab:workshopParticipants}
\Description{Table describing workshop Participants from Agency A and Agency B. There are 13 participants from Agency A, mostly social workers, and 14 participants from Agency B, a mix of youth workers, youth project workers, and counsellors.}
\end{table}

\section{Workshop Findings}
\label{appendix:workshopFindings}

We categorise our findings into three main themes: Difficulties faced by SSPs in their daily work, possible AI-assisted solutions, and potential risks of using AI systems.


\subsection{Difficulties in Social Service Practice}
\label{sec:stage1difficulties}

Echoing past sentiments \cite{singer_ai_2023, tiah_can_2024}, participants cited the \textit{need to document "anything and everything"} (TS3) as a major pain point. SSPs have to write systematic reports in different structured formats (e.g. bio-psychological scales (TS2), risk factor assessments (TS3)) and repack the same content for different stakeholders (TD2, TS3) like colleagues or other agencies, creating tedious duplicate work. Additionally, some SSPs struggle with putting their ideas into text, "knowing what to do" in practice but being unable to translate that well into writing (TD1). This results in a lack of ability to incorporate theoretical concepts in reports, something noted particularly by senior personnel (TD1, TD2). 

Participants also noted the \textit{challenges in case formulation}, the synthesising of information to craft assessments and justify a certain judgement (TS3). Piecing together disparate information across many pages of case notes is cognitively challenging and time-consuming (TD1). This is especially tricky during direct interaction with a client, where the non-linear process of information discovery (CC2) leaves workers constantly trying to organise their "all over the place" (CC5) thoughts to understand a case holistically before proceeding to ask the client the most pertinent questions. This may inadvertently cause them to miss certain key insights or red flags (TS2, TD1) only evident from looking at the gathered information as a whole.

\subsection{Possible Solutions}
\label{sec:stage1solutions}

Having elucidated these difficulties, participants floated numerous ideas for how AI could help with these issues. Many participants expressed a desire for a tool to help with \textit{manual labour}- for example, turning point-form notes into formal reports (CP1, CP2, CC1, TS2). TD1 shared how SSPs frequently used the "5Ps" framework\footnote{Presenting problem, Predisposing factors, Precipitating factors, Perpetuating factors, Protective factors} to organise case information into different, logically linked categories to aid subsequent analysis. This is one possible, largely menial and procedural task that an LLM could perform with high accuracy. Other forms of documentation mentioned, like the Data, Intervention, Assessment, Plan (DIAP), or Bio-Psychosocial-Spiritual (BPSS) formats, were also common report structures that could be automatically generated. 

More cognitively-intensive, \textit{mental labour} tasks were also considered, such as the generation of assessments (TS3, TS4, TS6, TS7). This requires fitting a client's information into theoretical frameworks to produce well-grounded assessments and analyses. AI tools could improve both the speed at which relevant information is synthesised and categorised, and the quality of the output through use of technical, industry-standard terms, something that is often lacking in many SSPs (TD1, TD2). Another critical task is intervention planning, where workers follow established models like Cognitive Behaviour Therapy (CBT) or Solution-Focused Brief Therapy (SFBT) and craft a plan for their clients based on the guiding steps in each model. An AI tool could allow the rapid generation of numerous possible interventions (CP1), leaving the user to pick and choose from the suggestions offered (TD1).

\subsection{Areas of Risk}
\label{sec:stage1risks}
In line with past work on the dangers of AI adoption, participants echoed certain concerns about the use of AI tools. Privacy of client data was unilaterally mentioned by all, a universal concern core to the social sector. Storage of personal client information (TS3, TP1) was a particular worry, and TS1 noted the possibility of workers entering sensitive information into a system by mistake. On the staff competency front, a common concern was the consequences of AI taking on an increasing part of the worker's job scopes. Specifically for analytical or ideative tasks, some seniors (TD1, TD2, TS1) were concerned about the loss in critical thinking skills of junior workers who might become overreliant on the tool to perform their work for them. Multiple participants also raised the possibility of inaccurate outputs from an AI system, particularly risky when less experienced workers fail to tell when the AI's output might be suboptimal and proceed to adopt its suggestions anyway. 

\section{Thematic Analysis}
\label{appendix:analysis}

To analyze the session transcripts, we performed qualitative thematic analysis \cite{braun2006using}. Qualitative thematic analysis is a method for identifying and analyzing key patterns and themes within interview and focus group data that affords a flexible and iterative coding process \cite{braun2006using}, making it particularly suited to the iterative nature of PD studies like this one. Due to the nascence of AI in the social service sector and the paucity of existing theoretical frameworks specific to the context of our study, we adopted an inductive, bottom-up approach of coding, which allowed us to uncover emergent themes in a data-driven manner and surface the latent thoughts, feelings and attitudes of SSPs in their own voice, rather than attempting to fit the data within a preexisting framework or being influenced by researchers' prior analytic preconceptions. 

Coding of the transcripts was carried out sequentially and iteratively. First, four researchers independently coded two transcripts and met to discuss and share the codes. Subsequently, two researchers then independently developed a codebook based on identified codes and met to combine the two into a unified codebook of themes and subthemes. The codebook was shared with the other two researchers alongside explanatory memos and used to code the remaining six transcripts, which were equally divided among all four researchers, with at least two researchers coding each transcript. The researchers then met two additional times to discuss the codes and resolve disagreements, until consensus was met. Finally, two researchers met in multiple rounds to collaboratively revise and refine the codes, as new themes were discovered and old ones were combined or retired. These themes form the final framework that informs our findings.

\end{document}